\documentclass[onecolumn,amsmath,amssymb,12pt,superscriptaddress,nofootinbib]{revtex4}
\pdfoutput=1

\usepackage[latin1]{inputenc}
\usepackage[english]{babel}
\usepackage{amssymb}
\usepackage{amsmath}
\usepackage{amsthm}
\usepackage[]{graphicx}
\usepackage[]{subfigure}
\usepackage{tensor}
\usepackage{color}
\usepackage{cancel}
\usepackage{setspace}
\usepackage{fancyhdr}
\usepackage[bookmarks,linktocpage, colorlinks=true, plainpages = false, citecolor = blue,  linkcolor=blue, urlcolor = blue, filecolor = blue]{hyperref} 
\newcommand{\ie}{{\it i.e.\ }}

\newcommand{\eg}{{\it e.g.\ }}

\begin{document}
\setcounter{page}{0}
\allowdisplaybreaks
\begin{titlepage}
\begin{flushright}
Imperial/TP/2019/KSS/02
\end{flushright} 
\vspace{.2in}

\title{A Safe Beginning for the Universe? 
\vspace{.3in}}

\author{Jean-Luc Lehners}
\email{jlehners@aei.mpg.de}
\affiliation{Max--Planck--Institute for Gravitational Physics (Albert--Einstein--Institute), 14476 Potsdam, Germany}
\author{K.\,S. Stelle}
\email{k.stelle@imperial.ac.uk}
\affiliation{Blackett Laboratory, Imperial College London, Prince Consort Road, London SW7 2AZ, U.K.}

\begin{abstract}
\vspace{.4in} \noindent 
When general relativity is augmented by quadratic gravity terms, it becomes a renormalisable theory of gravity. This theory may admit a non-Gaussian fixed point as envisaged in the asymptotic safety program, rendering the theory trustworthy to energies up to the Planck scale and even beyond. We show that requiring physical solutions to have a finite action imposes a strong selection on big-bang-type universes. More precisely we find that, in the approach to zero volume, both anisotropies and inhomogeneities are suppressed while the scale factor is required to undergo accelerated expansion. This provides initial conditions which are favourable to the onset of an inflationary phase while also providing a suitable starting point for the second law of thermodynamics in the spirit of the Weyl curvature hypothesis.
\end{abstract}
\maketitle

\end{titlepage}

\tableofcontents


\section{Introduction}

From observations of the cosmic background radiation we know that at an early epoch, the universe was highly isotropic and homogeneous, spatially flat to a good approximation, only containing small nearly scale-invariant and Gaussian density fluctuations \cite{Aghanim:2018eyx}. It is clear that this configuration of the universe was very special. But one may wonder whether it was the outcome of a special state in the very early universe, or whether it arose from a generic early state. That is to say, could the dynamics alone have been sufficient to explain the state of the universe at recombination, maybe because of a strong attractor? Or was there rather a selection rule that determined the initial state? A debate between these two viewpoints has been at the heart of cosmology for as long as people have thought about the beginning of the universe. 

With the advent of inflation \cite{Guth:1980zm,Linde:1981mu,Albrecht:1982wi}, many people started believing that dynamics alone would be sufficient\footnote{Though, interestingly, Starobinsky's early model was based on the premise that a de Sitter geometry is a special solution \cite{Starobinsky:1979ty}.}. More precisely, the idea was that with ``generic'' initial conditions (for instance with some sort of ``equipartition'' of energies at the Planck scale, as in chaotic models \cite{Linde:1983gd}) inflation would occur somewhere, and in that region the universe would expand so much that it quickly came to dominate over any surrounding region where inflation did not take hold. In all parts where inflation came to an end, the universe would then automatically be flat, isotropic and with the required density perturbations amplified from quantum fluctuations during inflation. However, observations put an upper limit on the Hubble rate during inflation, with the upper bound being $5$ orders of magnitude below the Planck scale \cite{Aghanim:2018eyx}. Since kinetic, gradient and potential energies scale very differently with the expansion of the universe, any kind of equipartition at the Planck scale would evolve to something far from the conditions required for inflation at the scale of the inflationary potential \cite{Ijjas:2013vea}, implying that inflation itself requires special initial conditions to get underway. Semi-classical quantisation of inflationary universes points to the same need for a suitable pre-inflationary phase \cite{Hofmann:2019dqu,DiTucci:2019xcr}. Meanwhile, one should bear in mind that we do not actually know whether an inflaton field, with a viable potential, actually exists. Dynamical mechanisms explaining the state of the universe are also central to cyclic models \cite{Steinhardt:2001st,Lehners:2008vx}, where an ekpyrotic contracting phase plays a role similar to that of inflation \cite{Khoury:2001wf}, while the bounce can act as an additional ``filter'' as envisaged in the phoenix universe \cite{Lehners:2008qe,Lehners:2009eg}. These models remain equally speculative, since we do not know if ekpyrotic matter exists and since the physics of the bounce remains work in progress \cite{Qiu:2011cy,Easson:2011zy,Koehn:2013upa,Ijjas:2016tpn,Dobre:2017pnt}. 

The opposite point of view is to find an explanation for a special initial state of the universe. This has been the approach pursued in the no-boundary proposal \cite{Hartle:1983ai}, the tunneling proposal \cite{Vilenkin:1983xq} and more recent ideas such as the CPT symmetric universe \cite{Gielen:2015uaa,Boyle:2018tzc}. So far the no-boundary and tunneling proposals have been considered in conjunction with inflation or ekpyrosis \cite{Battarra:2014xoa,Battarra:2014kga}, and a debate is still unfolding as to how to implement these theories in the best way \cite{Feldbrugge:2017fcc,DiazDorronsoro:2017hti,Feldbrugge:2017mbc,DiazDorronsoro:2018wro,Feldbrugge:2018gin,Halliwell:2018ejl,DiTucci:2019dji}. These proposals certainly remain promising candidate explanations for the state of the early universe, and besides the question of a precise mathematical implementation it will be interesting to see if they can lead to observationally distinguishable predictions. In the same camp of theories for the initial state resides Penrose's Weyl curvature hypothesis \cite{Penrose:1900mp,Penrose:1988mg}. Penrose noted that around the time of recombination the Weyl curvature was extremely low, while during gravitational collapse it increases (and diverges near the centres of black holes). Thus originated his suggestion that the Weyl curvature was zero (or as small as the uncertainty principle allows) at the big bang. If gravitational entropy is related to the Weyl curvature, this has the further implication of explaining the low initial entropy in the universe, thereby providing a suitable starting point for the second law of thermodynamics. But to date no concrete implementation of this proposal has been found.

In the present paper we want to investigate possible implications of quadratic gravity on the question of initial conditions\footnote{Quantum cosmological aspects of quadratic gravity have previously been studied in \cite{Hawking:2000bb}.}. It seems clear that general relativity (GR) will not be the final theory of gravity, as it must be quantised, but it happens to be non-renormalisable. However, upon inclusion of terms quadratic in curvature tensors, gravity does become renormalisable \cite{Stelle:1976gc}, and for this reason we will focus our attention on this theory. There is some evidence that quadratic gravity is in fact asymptotically safe \cite{Fradkin:1981iu,Avramidi:1985ki,Codello:2006in,Niedermaier:2009zz,Niedermaier:2010zz,Ohta:2013uca}, implying that it can be trusted up to Planck scale energies and even higher\footnote{We do not take a definite position here as to whether quadratic gravity is fully acceptable as an ultimate theory of gravity. It might be considered as an intermediate effective theory of gravity, valid to energies far above those for which Einstein's theory alone is applicable. A deeper theory may still be required to understand the microscopic nature of spacetime and gravitons.}. This makes it particularly appealing in terms of exploring the consequences for the big bang. It is here that we find a crucial difference between pure general relativity and the quadratic curvature terms. There exist many solutions of general relativity that contain an inhomogeneous, anisotropic spacelike singularity in their past. In fact, the singularity theorems of Penrose and Hawking tell us that this is generically the case \cite{Hawking:1969sw}. Despite the singularity many such solutions nevertheless have finite GR action and would thus not be questionable from the quantum point of view. However, we find that the quadratic gravity terms modify this conclusion: they generically cause the action to blow up if there are anisotropies and inhomogeneities present at the big bang. Since infinite action solutions cannot be part of a well-defined quantum theory of gravity, such spacetimes are filtered out. In this manner quadratic gravity selects a homogeneous and isotropic big bang, thus potentially explaining some of the most basic puzzles of the early universe. What is more, finiteness of the action requires the scale factor to undergo accelerated expansion, so that conditions favourable for a subsequent inflationary phase are obtained. 

Quantum quadratic gravity thus acts as a strong selection principle, as it requires a homogeneous and isotropic big bang. This implements the Weyl curvature hypothesis, and can therefore also explain the low entropy present at early times. Let us emphasise that the structure of the theory is already enough by itself to select a class of very special possible initial states, differing only in their early expansion rates. Moreover, it is conceivable that future work will show that even this rate is determined, \eg by the relevant renormalisation group behaviour of the cosmological constant. This would constitute a resolution of the debate mentioned at the beginning of this introduction, but without having to introduce further ad hoc principles: quadratic gravity combined with basic quantum principles may already be enough to explain why our universe had a safe beginning.


\section{Quadratic gravity}

The action that we will consider is that of ($4$-dimensional) general relativity, including a cosmological constant $\Lambda,$ augmented by terms that are quadratic in the Riemann tensor $R^\mu{}_{\nu\rho\sigma}$. As is well known, the Gauss-Bonnet identity implies that the particular combination $\sqrt{-g}\left(R^{\mu\nu\rho\sigma}R_{\mu\nu\rho\sigma}-4R^{\mu\nu}R_{\mu\nu}+R^2\right)$ forms a total derivative, so that we may restrict our attention to two independent quadratic gravity terms. We will take these to be the Ricci scalar squared, and the Weyl tensor $C^\mu{}_{\nu\rho\sigma}$ squared, which satisfies the relation $\int \sqrt{-g}\,C^{\mu\nu\rho\sigma}C_{\mu\nu\rho\sigma}=\int \sqrt{-g}\left(2R^{\mu\nu}R_{\mu\nu}-\frac{2}{3}R^2\right)$. Consequently our action can be written as 
\begin{align}
S & = \int d^4x \sqrt{-g} \left[ \frac{1}{\kappa^2} R - \Lambda -\frac{1}{2\sigma} C_{\mu\nu\rho\sigma}C^{\mu\nu\rho\sigma} + \frac{\omega}{3\sigma} R^2  \right]\,. \label{QuadGravity}
\end{align}
The gravitational coupling is given in terms of Newton's constant $G$ as $\kappa^2 = 16\pi G.$ In terms of dimensionless couplings we can write   $\frac{1}{\kappa^2} = \frac{\mu^2}{g_N}$ and $\Lambda = \lambda \mu^4,$ where $\mu$ is an energy scale and $g_N, \lambda$ are the dimensionless Newton coupling and cosmological constant respectively. The couplings of the quadratic terms, $\sigma$ and $\omega,$ are already dimensionless by definition. 

Pure general relativity is not renormalisable \cite{tHooft:1974toh,Christensen:1979iy,Goroff:1985th}, in the sense that an infinite number of counterterms, up to arbitrarily high powers in the curvature tensors, would be required to specify the high energy behaviour upon including loop corrections. This situation is dramatically improved if instead of pure general relativity one takes the action of quadratic gravity \eqref{QuadGravity} as the starting point \cite{Stelle:1976gc}. The $1/k^4$ momentum dependence of the propagator at high frequencies then renders this theory renormalisable, and does not require the inclusion of counterterms of yet higher orders in the curvature tensors. The evolution of the (dimensionless) couplings under a change of energy scale $\mu$ are given at one loop order by the following beta functions (calculated in a Euclidean setting \cite{Fradkin:1981iu,Avramidi:1985ki,Codello:2006in,Niedermaier:2009zz,Niedermaier:2010zz,Ohta:2013uca}),
\begin{align}
\mu \frac{d}{d\mu} g_N & = f_g(g_N,\lambda,\sigma,\omega)\,, \qquad \qquad \mu \frac{d}{d\mu} \lambda  =  f_\lambda(g_N,\lambda,\sigma,\omega)\,, \nonumber \\
\mu \frac{d}{d\mu} \sigma & = - \frac{133}{160\pi^2}\sigma^2\,,  \qquad \quad \,\, \mu \frac{d}{d\mu} \omega  = - \frac{25+1098 \omega + 200 \omega^2}{960\pi^2} \sigma\,.
\end{align}
Here $f_g$ and $f_\lambda$ are functions of all the couplings, but whose precise form we will not require. The beta functions for the higher-derivative couplings form a closed subsystem. Interestingly,  there is evidence for a non-trivial fixed point at $\sigma^\star=0,\omega^\star\approx -0.0228$ and at finite positive (but more strongly definition and scheme dependent) values $g_N^\star, \lambda^\star,$ with $\sigma$ approaching zero in inverse proportion to the logarithm of the energy scale, $\sigma \sim 1/\ln\mu$ \cite{Codello:2006in,Niedermaier:2009zz,Niedermaier:2010zz,Ohta:2013uca}\footnote{We should mention that other works have found evidence for another non-trivial fixed point at which the quadratic couplings would both take non-zero values \cite{Benedetti:2009rx}.}. The precise values of the asymptotic couplings will change once matter contributions are included, but our arguments will be independent of such refinements. The existence of a non-trivial fixed point is in line with the goals of the asymptotic safety programme initiated by Weinberg \cite{Weinberg:1980gg}. Note that at high energies only a specific linear combination of the quadratic terms, namely $C_{\mu\nu\rho\sigma}C^{\mu\nu\rho\sigma} - \frac{2 \omega^\star}{3}R^2,$ remains relevant to leading order in $\sigma$. (Most of the work on asymptotic safety has focussed on the full space of couplings of diffeomorphism invariant gravitational theories, \ie on general relativity extended by the infinite series of terms of higher orders in the Riemann curvature, see \cite{Reuter:1996cp,Percacci:2017fkn,Eichhorn:2018yfc} and references therein. Here we restrict to quadratic gravity because of its special renormalisability properties.)

To further understand the implications of these running couplings we can consider a metric fluctuation $h_{\mu\nu}$ around a reference metric $\bar{g}_{\mu\nu}.$  Since we are taking quadratic gravity as our starting point, we must rescale the metric fluctuations $h$ such that the kinetic term $\ln\mu (\partial^2 h)^2$ becomes canonical, i.e. we have to define $(\ln\mu)^{1/2} h \equiv \tilde{h}$. Then the higher-derivative kinetic term will be of the form $(\partial^2 \tilde{h})^2.$ Under such a re-scaling, the cubic interaction term contained in the Einstein-Hilbert term will be of the form $\mu^2/(\ln\mu)^{3/2} \tilde{h}^2\partial^2\tilde{h},$ demonstrating that the effective (dimensionful) gravitational coupling $\mu^2/(\ln\mu)^{3/2}$ blows up, with the consequence that gravity is strongly coupled at large energies (while the dimensionless coupling $1/(\ln\mu)^{3/2}$ tends to zero). The cubic interactions contained in the quadratic gravity terms go as $(\ln\mu)/(\ln\mu)^{3/2} = 1/(\ln\mu)^{1/2}$ and thus go to zero. In other words, the quadratic gravity terms become asymptotically free \cite{Fradkin:1981iu,Avramidi:1985ki}. 

The quadratic gravity terms have the effect of introducing additional degrees of freedom into the theory \cite{Stelle:1977ry}. Beyond massless gravitons, there are negative energy spin $2$ excitations of mass squared $\sigma/\kappa^2 \sim \mu^2/\ln\mu$ and spin $0$ excitations of mass squared $\frac{\sigma}{2\kappa^2 \omega} \sim \mu^2/\ln\mu.$ Note that the scalar degree of freedom is non-ghost \cite{Stelle:1977ry} but becomes tachyonic in the approach of the fixed point $\omega^\star < 0$ as $\mu \to \infty,$ although simultaneously the masses of the new degrees of freedom are pushed to infinity. The extent to which these additional excitations lead to instabilities remains a matter of debate, to which we have nothing new to add here. Here, we will consider quadratic gravity as an effective quantum theory of gravity, and we will explore consequences for the early universe that will be independent of this issue. 

The renormalisability and asymptotic safety of quadratic gravity imply that we can trust the theory up to arbitrarily high energies. If we now consider a transition amplitude, calculated for instance as a sum over paths weighted by the action, then we only obtain a relevant contribution to this sum if the action is well-defined and not divergent. This leads us to impose the following condition, motivated by basic quantum mechanics: all physical solutions of the theory should have \emph{finite action}. This criterion will be especially relevant near the big bang. Note that usually  one is not bothered by a divergence in the action near the big bang, as one would usually relegate this problem to new physics at the Planck scale. But here, since we can take the theory seriously as written down, we can also take solutions seriously up to arbitrarily high energies. Imposing the condition of finite action will turn out to be a strong selection principle. 

We should note that the relation between renormalisation scale and physical scales is not completely clear: often one can identify the energy scale with the ambient temperature, but in the early universe there may not have been any radiation at early stages, and even if so it may not have been in any sort of equilibrium. It may thus make more sense to identify the energy scale with the curvature scale stemming from the Ricci scalar, but more work is required in general to elucidate the relationship between renormalisation scales and cosmological evolution. What we will do in the present paper is the  following: we will assume that the metric is such that we approach zero volume as the time coordinate $t \to 0.$ Then at fixed (and arbitrarily large) scale $\mu$ we will impose the requirement that the action is convergent, in particular that the time integral in the action does not diverge as the initial time is taken to zero. In the same vein, we will assume that the spatial volume of the universe is finite. We are now in a position to explore the consequences of this criterion for early anisotropies and early inhomogeneities, which we will do in turn.


\section{Anisotropies}

A useful way to analyse anisotropies is to consider the Bianchi IX metric, which can also be regarded as a non-linear completion of a gravitational wave. This analysis will serve as a good indication for the fate of general anisotropies. The Bianchi IX metric can be written as
\begin{align}
ds_{IX}^2 = - dt^2 + \sum_m \left( \frac{l_m}{2} \right)^2 \sigma_m^2\,,
\end{align}
where $\sigma_1 = \sin\psi\, d\theta - \cos \psi \sin \theta\, d\varphi$, $\sigma_2 = \cos \psi\, d\theta + \sin \psi \sin \theta\, d \varphi$, and $\sigma_3 = -d\psi + \cos\theta\, d\varphi$ are differential forms on the three sphere with coordinate ranges $0 \leq \psi \leq 4 \pi$, $0 \leq \theta \leq \pi$, and $0 \leq \phi \leq 2 \pi.$ We may then rescale  
\begin{align}
l_1 = a \, e^{\frac{1}{2}\left(\beta_+ + \sqrt{3}\beta_-\right)}\,, \quad
l_2 = a \, e^{\frac{1}{2}\left(\beta_+ - \sqrt{3}\beta_-\right)}\,, \quad
l_3 = a \, e^{-\beta_+}\,,
\end{align}
such that $a$ represents the spatial volume while the $\beta$s quantify a change in the shape of spatial slices. When $\beta_- = \beta_+ = 0$ one recovers the isotropic case. The Einstein-Hilbert action in these coordinates is given by
\begin{align}
S_{EH} = \int d^4x \sqrt{-g} \frac{R}{2} = 2\pi^2 \int dt a \left( -3\dot{a}^2  + \frac{3}{4}a^2(\dot{\beta}^2_+ + \dot{\beta}^2_-) -  U(\beta_+, \beta_-)\right)\,,
\end{align}
where the anisotropy parameters evolve in the effective potential
\begin{align} \label{anisotropypotential}
U(\beta_+, \beta_-)  = - 2 \left( e^{ 2 \beta_+ } + e^{-\beta_+ - \sqrt{3}\beta_-} + e^{-\beta_+ + \sqrt{3}\beta_-} \right) + \left( e^{ -4 \beta_+ } + e^{2\beta_+ - 2\sqrt{3}\beta_-} + e^{2\beta_+ + 2\sqrt{3}\beta_-} \right)\,.
\end{align}
The Friedmann equation resulting from this action alone is given by (with $H=\dot{a}/a$)
\begin{align} \label{Friedman}
3H^2 = \frac{3}{4}\dot{\beta}^2_+ + \frac{3}{4}\dot{\beta}^2_-   + \frac{1}{a^2}U(\beta_+, \beta_-)\,,
\end{align}
and it can be used to simplify the on-shell action which is then given by the compact expression 
\begin{align}
S_{EH}^{on-shell} = \int d^4x \sqrt{-g} \frac{R}{2} = -4\pi^2 \int dt \, a  U(\beta_+, \beta_-)\,. \label{onshell}
\end{align}

The full action that we will consider includes the contributions due to quadratic gravity. The Weyl squared part is given by (up to total derivative terms)
\begin{align}
& - \int d^4x \sqrt{-g}\, C_{\mu\nu\rho\sigma}C^{\mu\nu\rho\sigma} \nonumber \\ = & \, 2\pi^2 \int dt   \Big{\{} 3 a^3 \left[\left( \frac{\ddot{a}}{a}+H^2 \right) (\dot\beta_{-}^2 + \dot\beta_{+}^2) -\ddot\beta_-^2 - \ddot\beta_+^2 -(\dot\beta_{-}^2 + \dot\beta_{+}^2)^2\right]  \nonumber \\ &  \, + 4a \left[ -(\ddot\beta_- + 3H \dot\beta_-) U_{,\beta_-} -(\ddot\beta_+ + 3H \dot\beta_+) U_{,\beta_+} -\left(2\frac{\ddot{a}}{a} + \dot\beta_-^2 + \dot\beta_+^2\right)U\right] \nonumber \\ &  \, + \frac{64}{3a} \left( - e^{-8\beta_+} + e^{-5\beta_+ -\sqrt{3}\beta_-} + e^{-5\beta_+ + \sqrt{3}\beta_-} - e^{-2 \beta_+} + e^{\beta_+ -3\sqrt{3} \beta_-} + e^{\beta_+ + 3 \sqrt{3} \beta_-}  \right.\nonumber \\ &  \left. \qquad \quad- e^{\beta_+ - \sqrt{3}\beta_-} - e^{\beta_+ + \sqrt{3}\beta_-} - e^{4\beta_+ -4\sqrt{3} \beta_-} - e^{4\beta_+ + 4 \sqrt{3}\beta_-}+ e^{4\beta_+ -2\sqrt{3} \beta_-} + e^{4\beta_+ +2\sqrt{3}\beta_-}\right) \Big\} \,.\label{actionweylsquared}
\end{align}
The anisotropy potential $U$ was defined in \eqref{anisotropypotential}. Meanwhile, the $R^2$ part is 
\begin{align}
  \int d^4x \sqrt{-g} R^2
=   \, 2\pi^2 \int dt  a^3 \left[ 6 \frac{\ddot{a}}{a} + 6\frac{\dot{a}^2}{a^2}  + \frac{3}{2}(\dot{\beta}^2_+ + \dot{\beta}^2_-) -  \frac{2}{a^2}U(\beta_+, \beta_-)\right]^2\,. \label{actionRsquared}
\end{align}

\begin{figure}[h]
\includegraphics[width=0.45\textwidth]{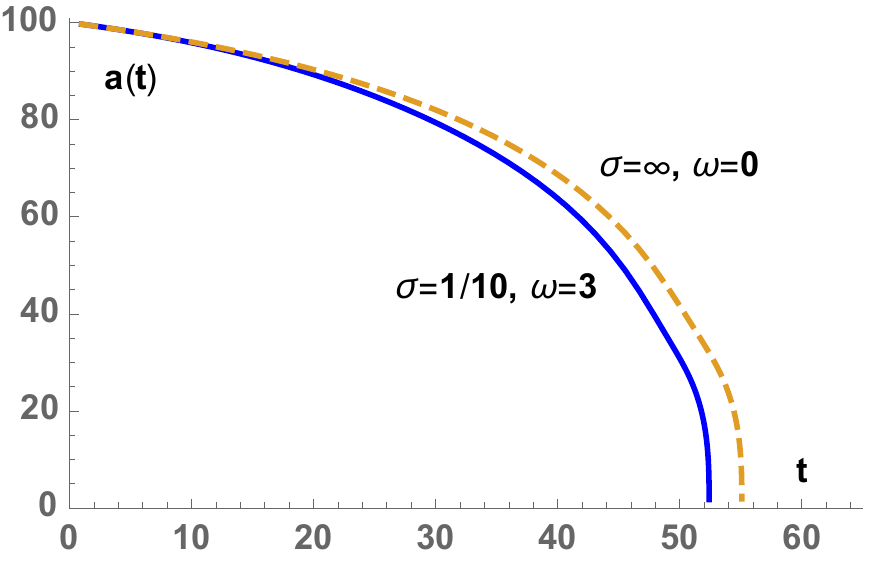}
\includegraphics[width=0.45\textwidth]{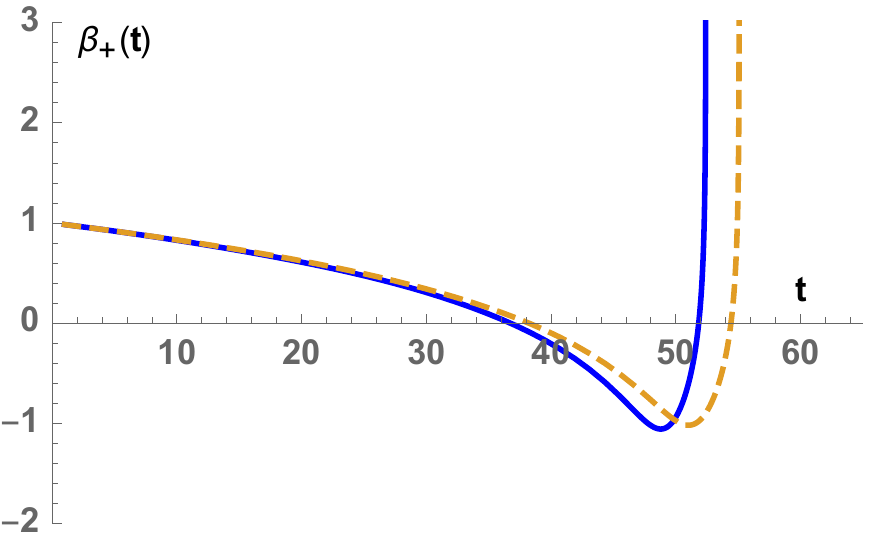}
\caption{Plots of the scale factor (left) and the anisotropy (right) as a function of time, for a typical example of the approach to a singularity. The dashed curves show the solution in the absence of quadratic gravity terms, where one can see that after one reflection of the anisotropy parameter, a singularity is reached at a finite time. When quadratic curvature terms are added, the only effect is to accelerate the occurrence of the crunch, as seen in the continuous curves. The initial conditions used here for the two-derivative theory are: $a(t=0)=100\,, \dot{a}(0)=-1/3\,, \beta_+(0)=1.$ For the case with higher-derivative terms added, we fixed the additional initial data by using the two-derivative solution, and we checked that the constraint remains satisfied to sufficient accuracy.} \label{fig:example}
\end{figure}

The solutions to the two-derivative equations of motion, starting from ``generic'' conditions at some time and evolving back towards the big bang, correspond to the Belinsky-Khalatnikov-Lifshitz (BKL)/mixmaster chaotic solutions \cite{Belinsky:1970ew,Misner:1969hg}, in which the universe evolves increasingly locally, with time derivatives being much more important than spatial derivatives, and contracting anisotropically. The solution then jumps between Kasner epochs, during which the anisotropy parameters evolve logarithmically, $\beta_\pm \sim \ln(t)$ and the universe contracts as $a(t) \sim t^{1/3}.$ We may wonder if this behaviour is significantly altered when the four-derivative terms are added to the action. To investigate this question, we may start by analysing the  equations of motion for small anisotropies, here retaining only the leading terms arising from the Weyl squared action. These are then given by 
\begin{align}
& 0  = \frac{1}{\kappa^2}\left( \ddot\beta_+ +3H\dot\beta_+ + 8 \frac{\beta_+}{a^2}\right) \nonumber \\ & + \frac{1}{\sigma}\left(  \beta_+^{(4)} + 6H\beta_+^{(3)} +\frac{4  \ddot{a} \ddot\beta_+}{a} +\frac{7  \dot{a}^2 \ddot\beta_+}{a^2} +\frac{20  \ddot\beta_+}{a^2} +\frac{4  \dot{a} \ddot{a} \dot\beta_+}{a^2}+\frac{ a^{(3)} \dot\beta_+}{a} +\frac{  \dot{a}^3 \dot\beta_+}{a^3}+\frac{20 \dot{a} \dot\beta_+}{a^3}   + \frac{64 \beta_+}{a^4} \right)\,, \label{linearbeta} \\
& 0  = \frac{1}{\kappa^2} \left(2\frac{\ddot{a}}{a} +H^2 +\frac{3}{4} \dot\beta_+^2 +\frac{1}{a^2}\right) + \frac{1}{\sigma}\left(\frac{1}{2}  \beta_+^{(3)}\dot\beta_+ -\frac{1}{4}   \ddot\beta_+^2 + H \dot\beta_+ \ddot\beta_+ +\frac{  \ddot{a} \dot\beta_+^2}{2a}+\frac{  \dot{a}^2 \dot\beta_+^2}{4 a^2}+\frac{5  \dot\beta_+^2}{a^2}  \right)\,, \label{lineara}
\end{align}
while the constraint reads
\begin{align}
& 0  = \frac{1}{\kappa^2}\left(3 H^2+\frac{3  }{a^2}-\frac{6   \beta_+^2}{a^2} -\frac{3}{4}   \dot\beta_+^2 \right) \nonumber \\ & + \frac{1}{\sigma}\left(-\frac{3}{2}  \beta_+^{(3)} \dot\beta_+ +\frac{3}{4}  \ddot\beta_+^2 -3  H \dot\beta_+ \ddot\beta_+ -\frac{3  \ddot{a} \dot\beta_+^2}{2a}-\frac{3  \dot{a}^2 \dot\beta_+^2}{4 a^2}-\frac{15  \dot\beta_+^2}{a^2}-\frac{48  \beta_+^2}{a^4}\right)\,.
\end{align}
At the level of the two-derivative theory we know that the relevant solution is the one with $a \propto t^{1/3}$ and $\dot\beta_+ \propto 1/t.$ So first we can try to solve around this solution, \ie we assume that $a \propto t^{1/3}$ and solve for the time dependence of $\beta_+$ using only the dominant terms in \eqref{linearbeta}. This yields four solutions $\beta_+ \propto t^b$ with $b \in \{0, \frac{2}{3}, \frac{4}{3}, 2 \}.$ All of these are decaying solutions as $t \to 0,$ except for $b=0$ which actually corresponds to the original $\dot\beta_+ \propto 1/t$ solution with $\beta_+$ growing logarithmically. The subleading terms will induce corrections. However, this shows that the higher-derivative terms do not drastically change the expected BKL behaviour in the approach to a spacelike singularity (there may however exist isolated islands of stability, as suggested by \cite{Barrow:2007pm}). Numerical studies of the full equations of motion confirm this expectation, and show that typically the solutions crunch faster when the higher-derivative terms are added -- an example is shown in Fig.\ \ref{fig:example}.

As we just saw, when the higher derivative terms are added the anisotropies remain large, and the behaviour remains chaotic, near the big bang. In this sense the higher order terms do not help in understanding the early state of the universe. However a theory of initial conditions must restrict the allowed solutions -- otherwise it has no explanatory power. The way this is achieved here is via a consideration of the action: our premise, based on quantum theory, will be that a solution with infinite action does not contribute to physical transition amplitudes. In this respect there is an interesting difference between the pure Einstein-Hilbert theory, and the theory with the quadratic curvature terms added in. 

A special but illustrative example (without reflections off the anisotropy walls) is the solution along one of the symmetry axes of the Bianchi IX spacetime, \ie a solution with $\beta_-=0.$ The anisotropy potential reduces to $U(\beta_+,0)=e^{-4\beta_+}-4e^{-\beta_+},$ and the asymptotic solution is given by
\begin{align}
a(t)=a_0 t^{1/3}\,\qquad e^{\beta_+}=b_+ t^{-2/3}
\end{align}
As $t \to 0$ the anisotropy diverges, \ie $\beta_+ \to \infty.$  The on-shell Einstein-Hilbert action \eqref{onshell} can then be approximated by
\begin{align}
S_{EH}^{on-shell} = - 4\pi^2 \int_{t_0}^{t_1} dt \, a U(\beta_+, 0) \approx 16\pi^2 \frac{a_0}{b_+}\int_{t_0}^{t_1} dt \, t\,,
\end{align}
where the approximate expression holds for small $t.$ We can see that the action will be finite as the lower bound of integration tends to zero, $t_0 \to 0.$ Thus, within the pure Einstein-Hilbert theory, there is no obstruction to considering a highly anisotropic solution of this kind. But now we can also evaluate the action when the quadratic curvature terms are included. We may, for instance, look at the Weyl squared term. With $\beta_-=0,$ it reduces to
\begin{align}
& - \int d^4x \sqrt{-g}\, C_{\mu\nu\rho\sigma}C^{\mu\nu\rho\sigma} \qquad \qquad (\beta_-=0)\nonumber \\ = & \, 2\pi^2 \int dt   \Big{\{} 3 a^3 \left[\left( \frac{\ddot{a}}{a}+H^2 \right) \dot\beta_{+}^2  - \ddot\beta_+^2 -\dot\beta_{+}^4 \right]   \, + 4a \left[ -(\ddot\beta_+ + 3H \dot\beta_+) U_{,\beta_+} -\left(2\frac{\ddot{a}}{a}  + \dot\beta_+^2\right)U\right] \nonumber \\ &  \, \qquad \qquad+ \frac{64}{3a} \left( - e^{-8\beta_+} + 2e^{-5\beta_+ }  - e^{-2 \beta_+}   \right) \Big\}\,. \label{actionweylsquaredbetaplus}
\end{align}
Near $t=0,$ for our example this action is approximately given by
\begin{align}
- \int d^4x \sqrt{-g}\, C_{\mu\nu\rho\sigma}C^{\mu\nu\rho\sigma} \approx 2\pi^2 a_0^3 \int_{t_0}^{t_1} dt \, \frac{56}{9 t^3} + \cdots\,,
\end{align}
where we have dropped subdominant terms. This clearly diverges as $t_0 \to 0,$ hence on this basis we would exclude this solution. Thus we have an example of a solution which has finite Einstein-Hilbert action, but infinite Weyl squared action and diverging Weyl tensor. The question now is how general this behaviour is. 

Based on the BKL analysis, we know that close to a spacelike singularity the generic solution is locally of Bianchi IX form, and for short periods of time this is well approximated by Kasner solutions. Thus we may try an ansatz 
\begin{align}
a \propto t^s\,\qquad e^{\sqrt{3} \beta_-} \propto t^m\,\qquad e^{\beta_+} \propto t^p\,.
\end{align}
The chaotic mixmaster solutions imply that most of the time, \ie in between reflections off the potential walls, we will have $s=1/3$ and also $m$ and $p$ will be related to each other. But in our context it makes sense to keep these coefficients general. As explained in the previous section, the Einstein-Hilbert term remains important at high energies. Near $t=0,$ the various terms in the Einstein-Hilbert action then scale as powers of time with the exponents
\begin{align}
& 3s-1\,, \quad \nonumber \\ &  s+2p+1\,, \quad s-4p+1\,,\quad \nonumber \\ & s-p-m+1\,,\quad s-p+m+1\,,\quad s+2p-2m+1\,,\quad s+2p+2m+1
\end{align}
where we are showing the integrated version (\ie we have performed the integral over $t$ -- for instance, the first term $\int a\dot{a}^2$ leads to $\int t^{3s-2} \propto t^{3s-1}$). In the absence of cancellations between terms (which we generically cannot expect to arise), we would want all of these exponents to be positive, implying the conditions
\begin{align}
s>\frac{1}{3}\,,\quad -\frac{1}{2}(1+s)<p<\frac{1}{4}(1+s)\,,\quad -\frac{1}{2}(1+s)<p \pm m< 1+s\,.
\end{align}
These conditions can easily be satisfied -- note that they allow for negative values of $p$ and $m,$ and consequently to a blowing up of anisotropy (and of Weyl curvature) near $t=0$. Convergence of the terms involving the scale factor alone simply leads to the requirement $s>1/3.$ 

Meanwhile, with the above ansatz, the Weyl squared and $R^2$ actions lead to the scalings (again barring cancellations between terms)
\begin{align}
& 3s-3\,,\quad s-1-p-m\,,\quad s-1-p+m\,,\quad s-1+2p+2m\,,\quad s-1+2p-2m \nonumber \\
& s-1+2p \,,\quad s-1-4p\,,\quad 1-s-8p\,,\quad 1-s-5p-m\,,\quad 1-s-5p+m\,,\quad 1-s-2p \nonumber \\
& 1-s+p+3m\,,\quad 1-s+p-3m\,,\quad 1-s+p+m\,,\quad 1-s+p-m\,,\quad \nonumber \\ 
&1-s+4p+4m\,,\quad 1-s+4p-4m\,,\quad 1-s+4p+2m\,,\quad 1-s+4p-2m
\end{align} 
Requiring these exponents to be positive leads to the conditions
\begin{align}
& s>1\,, \quad -\frac{1}{2}(s-1) < p < \frac{1}{4}(s-1) \,, \quad -\frac{1}{2}(1-s) < p < \frac{1}{8}(1-s) \nonumber \\
& p+m > s-1\,,\quad p+m < s-1\,,\quad p-m < s-1\,,\quad p-m > s-1\,,
\end{align}
which are clearly mutually exclusive. We can only obtain a finite action if the anisotropy parameters $\beta_\pm$ become zero near $t=0.$ As the universe expands the anisotropies will remain small at the level of the background (if an inflationary phase ensues, fluctuations will be produced by further quantum effects).  Thus the imposition of finite action suppresses anisotropies quite generally at the earliest times.


\section{Inhomogeneities}

What about spatially dependent (inhomogeneous) perturbations? Are they also suppressed, or can there be large variations in space? To answer this question it is instructive to consider a metric of the form
\begin{align}
ds^2 = -dt^2 + \frac{A^{\prime 2}}{F^2} dr^2 + A^2 (d\theta^2 + \sin^2\theta d\phi^2)\,,
\end{align}
which belongs to the Lema\^{i}tre-Tolman-Bondi class \cite{Tolman:1934za,Bondi:1947fta}. Here the scale factor depends not only on time, but also on the radial direction $r$, that is to say $A=A(t,r).$ A prime denotes a derivative w.r.t.\ $r.$ Meanwhile $F=F(r)$ is a function of $r$ only that describes the inhomogeneity in the $r$ direction. A spherical symmetry in the two remaining spatial directions is retained for simplicity. Relaxing this assumption would only strengthen the arguments below. When $F(r)=1$ and $A$ factorises into a time-dependent and an $r$-dependent function, then after a field redefinition this metric reduces to the standard flat Robertson-Walker form.  

The Einstein-Hilbert action is given by 
\begin{align}
\int d^4x \sqrt{-g} \, \frac{R}{2} = 4\pi \int dt \, dr \, \frac{A^\prime}{F}\left(1 - F^2 + \dot{A}^2 -2 \frac{AFF^\prime}{A^\prime} +2\frac{A\dot{A}\dot{A}^\prime}{A^\prime}+2A\ddot{A} + \frac{A^2}{A^\prime}\ddot{A}^\prime \right)\,.
\end{align}
As for the quadratic gravity terms, the $R^2$ action is given by 
\begin{align}
\int d^4x \sqrt{-g} \, R^2 = 16\pi \int dt \, dr \, \frac{A^\prime}{A^2 F}\left(1 - F^2 + \dot{A}^2 -2 \frac{AFF^\prime}{A^\prime} +2\frac{A\dot{A}\dot{A}^\prime}{A^\prime}+2A\ddot{A} + \frac{A^2}{A^\prime}\ddot{A}^\prime \right)^2\,,
\end{align}
while the Weyl squared action is
\begin{align}
\int d^4x \sqrt{-g} \, C_{\mu\nu\rho\sigma}C^{\mu\nu\rho\sigma} =  \, \frac{16\pi}{3} \int dt \, dr \,\frac{A^\prime}{A^2 F}\left(1 - F^2 + \dot{A}^2 + \frac{AFF^\prime}{A^\prime} - \frac{A\dot{A}\dot{A}^\prime}{A^\prime}- A\ddot{A} + \frac{A^2}{A^\prime}\ddot{A}^\prime \right)^2\,.
\end{align}
We are interested in the conditions under which the action is finite when integrated from zero spatial volume onwards, \ie we would like to see under what conditions the action integral diverges in the approach to the big bang. For this we will again look at the scalings of the various terms contributing to the action. As above, we will make the ansatz that the scale factor has a power-law time dependence in the approach to $A=0,$ \ie we will posit $A(t,r) \sim t^s$ near $t=0$,\  with $s$ being a positive number.  The Einstein-Hilbert action then has two different types of terms: those related to the time evolution  of the scale factor, such as $\int dt \, \frac{\dot{A}^2A^\prime}{F} \sim \int dt\, t^{3s-2} \sim t^{3s-1}$ and which require $s>1/3$ for convergence, and those related to the inhomogeneity, such as $\int dt \, A^\prime F \sim \int dt\, t^{s} \sim t^{s+1}.$ These latter terms are always convergent, since $s>0$ by assumption. Thus we can see that inhomogeneities lead to no divergence of the Einstein-Hilbert action, and would be perfectly acceptable solutions in the context of pure Einstein gravity. 

Now we can repeat this analysis for the quadratic gravity terms. The same line of reasoning shows that both the $R^2$ and the Weyl squared actions lead to temporal scalings of the form
\begin{align}
\int dt \, t^{3s-4}, \int dt \, t^{s-2}, \int dt \, t^{-s}\,,
\end{align}
where the last term comes from the pure inhomogeneity terms like $\int \frac{A^\prime F^4}{A^2 F}.$ Convergence near $t=0$ then requires 
\begin{align}
s>1\,,\quad s < 1\,.
\end{align}
The second requirement, which comes from the inhomogeneity contributions, is in clear conflict with the first one. This shows that near $t=0$ the inhomogeneity must be damped, $F(r) \to 1,$ in order for the action to remain finite. In this manner only initially homogeneous universes are allowed, with the scale factor undergoing accelerated expansion, $A \sim t^s$ with $s>1.$


\section{Discussion}

Quadratic gravity is not only in complete agreement with gravity experiments, but it is a renormalisable theory, which may in fact be asymptotically safe and thus trustworthy up to high energies. Taking this theory seriously then has the immediate and surprising consequence of explaining the special nature of our universe at the earliest times: the big bang is required to have been homogeneous and isotropic. Thus basic aspects of the large scale properties of our universe would be explained by the quantum requirement that physical amplitudes must have finite action. At the same time, such a safe beginning of the universe has vanishingly small Weyl curvature, and can thus explain the low initial entropy required for the unfolding of the second law of thermodynamics.  

Our arguments are simple and direct, but make use of a number of assumptions that deserve clarification in future work. Especially the relation between energy scale and curvature needs to be better understood, not just in the present context but more generally in early universe cosmology. An improvement on this front would also allow one to include the running of couplings into the equations of motion, and thus determine more precisely the evolution after the big bang: how long would a subsequent cosmological-constant-driven inflationary phase last (see also \cite{Weinberg:2009wa})? And could the energy scale be such that the observed amplitude of fluctuations in the cosmic background could be explained? Is the running of the $R^2$ term such that it can grow in relative importance compared to the Einstein-Hilbert term (and obtain the correct sign in its coupling) such that an inflationary phase of Starobinsky type could take place? Moreover, if the Higgs scalar is added, what are the properties of its potential at high energies, and would it influence the inflationary dynamics implied by the cosmological constant or $R^2$ terms? Most importantly: does the scenario described here lead to any distinguishable observational consequences? This will of course be crucial in assessing the ultimate validity of our proposal.

\acknowledgments

We would like to thank Andrei Barvinsky, Steffen Gielen, Caroline Jonas and Roberto Percacci for helpful discussions and correspondence. JLL gratefully acknowledges the support of the European Research Council in the form of the ERC Consolidator Grant CoG 772295 ``Qosmology''.  JLL would like to thank Imperial College for hospitality. The work of KSS was supported in part by the STFC under Consolidated Grant ST/P000762/1. KSS would like to thank the Albert Einstein Institute for hospitality on several occasions during the course of the work.

\bibliographystyle{utphys}
\bibliography{SafeBeginning}

\end{document}